\documentstyle[multicol,aps,epsf,epsfig]{revtex}
\newcommand{\lb}[1]{\label{#1}}
\newcommand{\bc}{\begin{center}}
\newcommand{\ec}{\end{center}}
\newcommand{\be}{\begin{equation}}
\newcommand{\ee}{\end{equation}}
\newcommand{\bea}{\begin{eqnarray}}
\newcommand{\eea}{\end{eqnarray}}
\newcommand{\ba}[1]{\begin{array}{#1}}
\newcommand{\ea}{\end{array}}

\newcommand{\bt}[1]{\begin{table}[ht]\centering\begin{tabular}{#1}}
\newcommand{\et}[1]{\end{tabular}\caption{\small#1}\end{table}}
\newcommand{\mod}{\,{\mathrm{mod}}\,}

\begin{document}
\title{Phase Diagrams of the Harper Map and the Golden Staircase}
\author{P. Castelo Ferreira$^1$\cite{pedro}, F. P. Mancini$^1$\cite{francesco} and M. H. R. Tragtenberg$^{1,2}$\cite{marcelo}}
\address{$^1$Department of Physics, University of Oxford, 1 Keble
Road, Oxford, OX1 3NP, UK\\
$^2$Depto. de F\'{\i}sica, Univ. Fed. de Santa Catarina,
Florian\'opolis, SC, Brazil, CEP 88040-900}
\date{\today}
\maketitle

\begin{abstract}
We present phase diagrams of the Harper map, which is equivalent to
 the problem of Bloch electrons in a uniform magnetic field 
(Azbel-Hofstadter model). We consider the cases where
 the magnetic flux $\omega$ assumes either the continued fraction
approximations towards the golden mean or the golden mean itself.
 The phase diagrams for  rational values of $\omega$ show a finite
number of Arnol'd tongues of localized electronic states with rational
winding numbers and regions of extended phases in between them.
 For the particular case of $\omega = \frac{\sqrt{5}-1}{2}$, we find
an infinite number of Arnol'd tongues of localized phases with
extended phases in between. In this case, the study of the winding
 number gives rise to a Golden Staircase, where the
plateaux represent localized phases with winding numbers equal to sums
of powers of the golden mean. We also present evidence of
the existence of an infinite number of strange nonchaotic attractors
for $\epsilon=1$ in points analogous to critical points in the
 pressure-temperature phase diagram of the water. [OUTP-00-07S]\\
{\footnotesize\noindent
Pacs: 03.65.Sq,05.45.-a,71.30.+h\ \ Keywords: Azbel-Hofstadter, Harper equation, Harper map,
phase transitions, strange nonchaotic attractors, fractal, golden staircase}
\end{abstract}
\begin{multicols}{2}

The Schr\"odinger equation for one electron in a two-dimension 
periodic lattice on a 
uniform magnetic field, in the tight binding approximation and Landau
gauge, is known as the  Azbel-Hofstadter problem~\cite{Azbel,Hofstadter}.
The discretized version of this model is the Harper equation~\cite{Harper} 
\be
\psi_{n+1}+\psi_{n-1}+2\epsilon\cos [2\pi(\phi_0+\omega n)]\psi_n=E\psi_n
\lb{schro}
\ee 
where $\psi_n$ is the discretized wave function, $\epsilon$ is the
lattice potential strength and
$\omega$ is the magnetic flux per unit cell in units of the quantum
flux (for reviews see~\cite{Souillard,Sokoloff}).
Under the transformation of variables $x_n=\psi_{n-1}/\psi_n$ we
obtain the Harper map~\cite{Ket-Sat}
\be
x_{n+1}=\frac{-1}{x_n-E+2\epsilon\cos(2\pi\phi_n)},
\lb{harper}
\ee
where $\phi_n=(\phi_0+\omega n) \mod 1$. We take $\phi \in [0,1[$ due
to the cosine periodicity.

The Lyapunov exponent related to the $\phi$ dynamics is $0$ and the
one concerning the variable $x$ is given by
\be
\lambda=\lim_{N\to\infty}\frac{1}{N}\sum^{N}_{n=0}y_n,
\ee
where 
\be
y_n=\log\frac{\partial x_{n+1}}{\partial x_n}=\log x_{n+1}^2.
\ee
Prasad {\em et al.}~\cite{Prasad} have studied in detail the
Localized-Extended transition for $\omega =  \omega^* =
(\sqrt{5}-1)/2$ (golden mean),
$\epsilon=1$ and $E=0, \pm 2.597\ldots$ 
(center and band edges of the spectrum). 
They found interesting connections 
between strange nonchaotic attractors (SNA)~\cite{Grebogi}
and localized-extended transition in the electronic wave function.
We extend their analysis for all values of E and other 
typical rational values of $\omega$.

Our study is based on the analysis of the Lyapunov exponent $\lambda$
as a function of the parameters $\epsilon, E$ and $\omega$.
Aubry and Andr\'e~\cite{aa} proved that the Lyapunov exponent
$\lambda$ is proportional to the localization length $\gamma$:
\be
\lambda=\lim_{N\to\infty}\frac{1}{N}\log\left(\frac{\psi_0}{\psi_N}\right)^2=-2\gamma.
\ee
For the Harper map $\lambda\le 0$. If $\lambda=0$, the phase 
is Extended (E), but for $\lambda<0$ the phase is
Localized (L).

All the characteristics discussed here are numerical results.
We start our analysis of the Harper map by discussing the phase
diagrams $\epsilon \times E$ for successive approximations 
$\omega_m=F_{m-1}/F_m$ to the golden mean, where $ F_0=F_1=1$ and
$F_m = F_{m-1}+F_{m-2}$ (Fibonacci sequence).  In this paper we show
the results for $\omega = 5/8$ and $8/13$, which display the essential
features of the phase diagrams for other orders of approximation for
the golden mean. Then, we exhibit the phase diagram for $\omega=\omega^*$.

The $\epsilon \times E$ phase diagrams are symmetric by reflection in
relation to the axes $E=0$ and $\epsilon=0$. Then, we will refer only
to their first quadrant. For a given $\omega_m$, there are
$F_{m-1}(F_{m-1}-1)$ connected regions of L(E) phases.
The particular stability limits of
each phase depend only on the value of the initial condition for
the variable $\phi$, namely $\phi_0$.  This behavior gives rise to  
co-stability regions, where more than one configuration is stable.
The stability
limits for different $\phi_0$ are invariant under the
transformation $\phi_0 \rightarrow \phi_0 + 1/F_m$.

\begin{figure}[ht]
\narrowtext
\begin{picture}(73,55)(0,0)
 \put(-13,-5){\epsfxsize=3.8in\epsffile{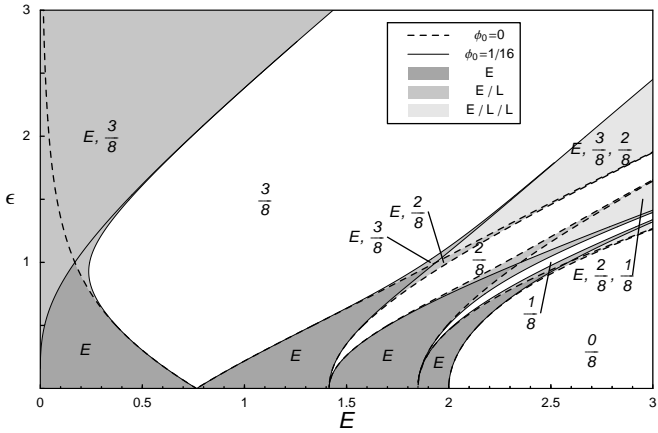}}
\end{picture}
\caption{Phase diagram for $\omega=5/8$. The labels represent the
winding numbers $\Omega$ of Localized phases. In the white regions there are
Localized phases, in lighter shaded regions there are co-stability of L
and E configurations and in the darkest shaded regions there are Extended phases.}
\label{fig:w58} 
\end{figure}

\begin{figure} 
\narrowtext
\begin{picture}(73,33)(0,0)
 \put(-5,-2){\epsfxsize=3.5in\epsffile{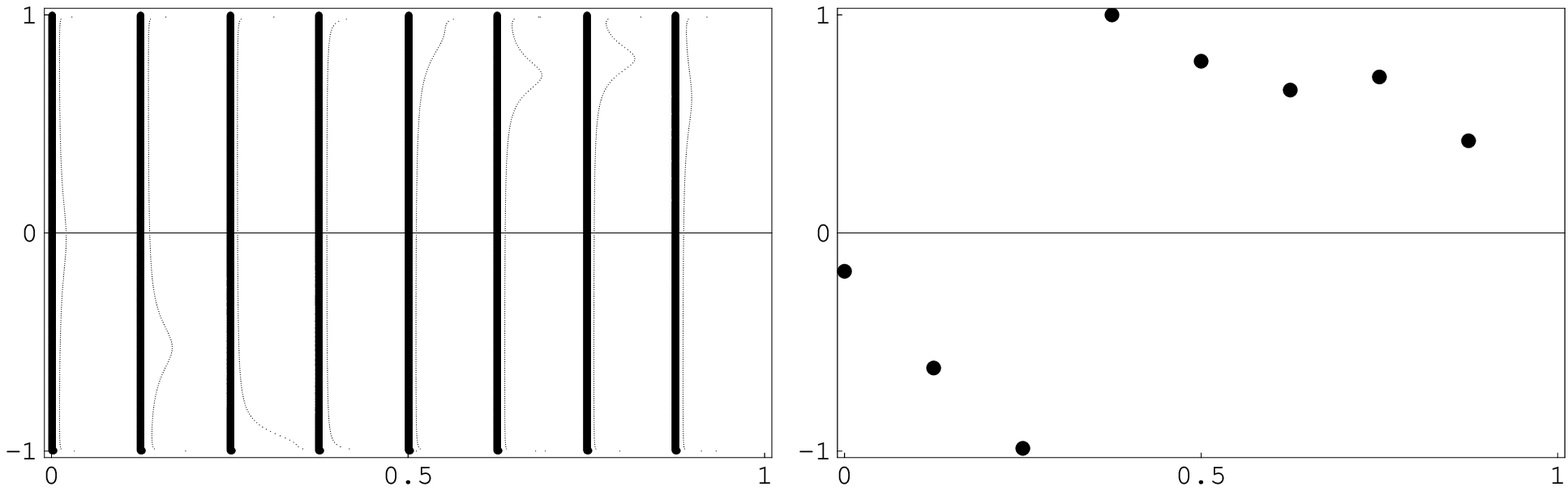}}
 \put(21,1){\footnotesize$\phi$}
 \put(60,1){\footnotesize$\phi$}
 \put(-1,11){\rotatebox{90}{\footnotesize$\tanh(x)$}}
 \put(21,28){$(a)$}
 \put(60,28){$(b)$}
\end{picture}
\caption{E and L configurations for $\omega = 5/8$ and $\epsilon
=0.5$. The E-L transition takes place at $E = 0.370773...$
(a)~Extended phase, $E=0.34$; the thick lines represent the 
configuration and the thin lines
the probability distribution of the iterations. (b)~Localized phase,
$E=0.39$; this attractor has winding number $\Omega=3/8$.}
\label{fig:LE58} 
\end{figure}

Fig. \ref{fig:w58} shows the phase diagram for $\omega = 5/8$. We observe
a {\em finite number of Arnol'd tongues}, a new feature in phase
diagrams, as far as we know. The white regions (L phases) are labelled
by winding numbers~\cite{winding} 
$\Omega = p/F_{m}$, where $p=0,1,2,\ldots,F_{m-2}$. In the
lighter shaded regions we can find either L or E configurations, 
{\em depending on the
value of $\phi_0$}. In the darkest shaded regions there are
E phases. The dashed curves represent the stability limits of the
Localized phases for $\phi_0 = 0$, whereas the continuous lines represent
these limits for $\phi_0 = 1/16$. 
In the $E=0$ axis there are two cases. For even $F_m$, there is
an E phase for $0 \le \epsilon \le \epsilon_1<1$; for $\epsilon_1 <
\epsilon < \epsilon_2$ there is an E/L
co-stability region. For $\epsilon>\epsilon_2$ there is a L phase.
For odd $F_m$, there is an E phase if $0 \le
\epsilon \le 1$ and a L phase for $\epsilon>1$. 
In the limit $m \rightarrow \infty$, $\epsilon_1$ and $\epsilon_2
\rightarrow 1$,and the behavior for odd and even $F_m$ become the same~\cite{Prasad}.
\vskip -.1 cm
The attractors of the Harper map corresponding to L phases, 
for a given $\omega_m$, are $F_m-cycles$, i.e., cycles with
period $F_m$. In the particular case of Fig.~\ref{fig:w58}, they are 8-cycles.
The configurations corresponding to Extended phases are
unidimensional. For $\omega = 5/8$, typical E and L configurations
 are represented in Fig.~\ref{fig:LE58}.
\vskip -.7 cm
In general, for a given $\omega_m=F_{m-1}/F_m$,  the configuration of an
Extended phase is made up by $F_m$ vertical lines in the plane
$\tanh(x)\times \phi$. 
The E-L transition occurs when the most visited
points in the unidimensional E phase turn to be a $F_m-cycle$ of the
Localized phase, as the energy and/or $\epsilon$ varies.

\begin{figure}
\narrowtext
\begin{picture}(73,45)(0,0)
 \put(-10,-46){\epsfxsize=4.6in\epsffile{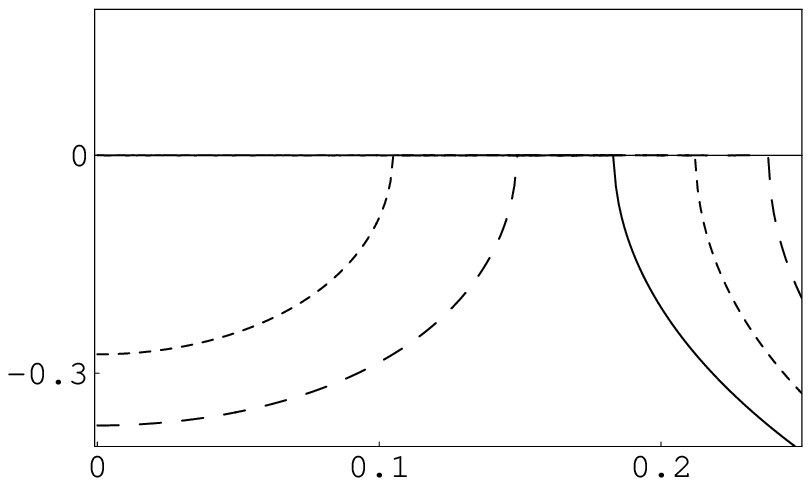}}
 \put(32,-51.5){\epsfxsize=5in\epsffile{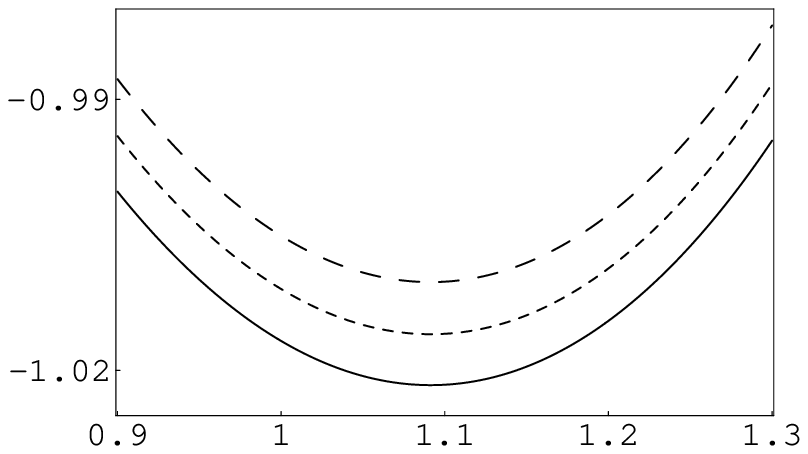}}
 \put(-3,17){\footnotesize$\lambda$}
 \put(42,17){\footnotesize$\lambda$}
 \put(18,3){\footnotesize$E$}
 \put(64,3){\footnotesize$E$}
 \put(18,26){$(a)$}
 \put(64,26){$(b)$}
\end{picture}
\caption{The Lyapunov exponent $\lambda$ vs. $E$
for $\omega=5/8$, $\epsilon=0.95$ and for $\phi_0=0$ (continuous
line), $\phi_0=1/32$ (dashed line),
$\phi_0=1/16$ (long dashed line). (a) shows co-stability between Extended
and Localized configurations; (b) shows co-stability between different
Localized attractors.}
\label{fig:cost}
\end{figure}

\begin{figure} 
\narrowtext
\begin{picture}(73,43)(0,0)
 \put(4,-2){\epsfxsize=2.6in\epsfysize=1.9in\epsffile{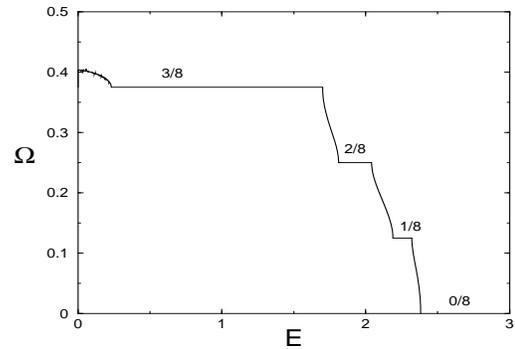}}
\end{picture}
\caption{Typical staircase for rational values of $\omega$,
representing the winding number $\Omega$ as a function of the
energy. Here, $\omega=5/8$, $\epsilon=0.8$ and $ \phi_0 = 5/8$. There
are four plateaux corresponding to the winding numbers 3/8, 2/8, 1/8
and 0/8. The winding number for $E=0$ is also 3/8. These plateaux are
connected through continuous lines,
representing the Extended configurations.}
\label{fig:wind58}
\end{figure}

The Localized phases within the white regions in Fig.~\ref{fig:w58} are
actually {\em co-stability regions}. There are many stable $F_m-cycles$
within each Arnol'd tongue, all of them with the same winding number.
Fig.~\ref{fig:cost} represents the Lyapunov
exponent $\lambda$ as a function of the energy, for three different
values of $\phi_0$. We see in
Fig.~\ref{fig:cost}a the co-stability of an Extended configuration with two
Localized attractors. In Fig.~\ref{fig:cost}b we see three stable
attractors, all with winding number $\Omega=3/8$.
\vskip -1.7 cm
The winding number changes with the energy. Fig.~\ref{fig:wind58} shows
the variation of $\Omega$ as a function of $E$ for $\omega=5/8$.
We see plateaux with winding numbers 3/8, 2/8,
1/8 and 0/8, which represent Localized phases.

\begin{figure} 
\narrowtext
\begin{picture}(73,55)(0,0)
 \put(-8,-42){\epsfxsize=6.5in\epsffile{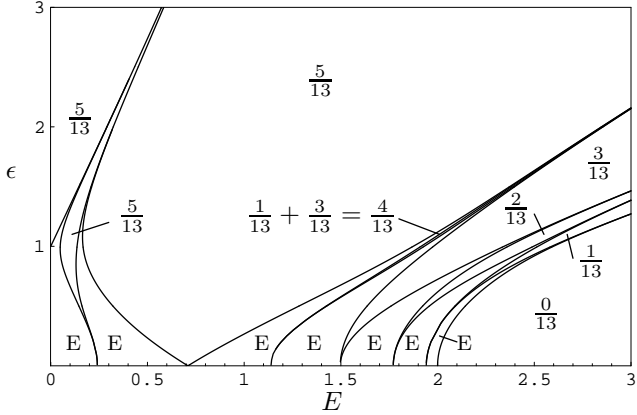}}
 \put(42,2){$E$}
 \put(0,33){$\epsilon$}
 \put(8,10){\footnotesize E}
 \put(13.5,10){\footnotesize E}
 \put(33,10){\footnotesize E}
 \put(40,10){\footnotesize E}
 \put(48,10){\footnotesize E}
 \put(53,10){\footnotesize E}
 \put(60,10){\footnotesize E}

 \put(8,40){$\frac{5}{13}$}
 \put(15,27){$\frac{5}{13}$}
 \put(40,45){$\frac{5}{13}$}
 \put(32,27){$\frac{1}{13}+\frac{3}{13}=\frac{4}{13}$}
 \put(77,34){$\frac{3}{13}$}
 \put(66,28){$\frac{2}{13}$}
 \put(76,21){$\frac{1}{13}$}
 \put(70,14){$\frac{0}{13}$}

\end{picture}
\caption{ Stability limits of the Localized and Extended phases for
$\omega=8/13$, $\phi_0=0$. }
\label{fig:w813} 
\end{figure}

The next approximant of $\omega$ towards $\omega^*$ is $8/13$. The number
of Arnol'd tongues, as well as connected Extended
regions have increased. The stability limits for $\phi_0=0$ are plotted in
Fig.~\ref{fig:w813}, but the full phase diagram must take into account
all initial conditions. Nevertheless, the structure of tongues with
pure Localized phases surrounded by L-E co-stability regions 
is the same as in Fig.~\ref{fig:w58}. The winding numbers
are labelling the Localized regions, made up by 13-cycles.

\begin{figure} 
\narrowtext
\begin{picture}(73,58)(0,0)
 \put(0,3){\epsfxsize=3.2in\epsfysize=2.3in\epsffile{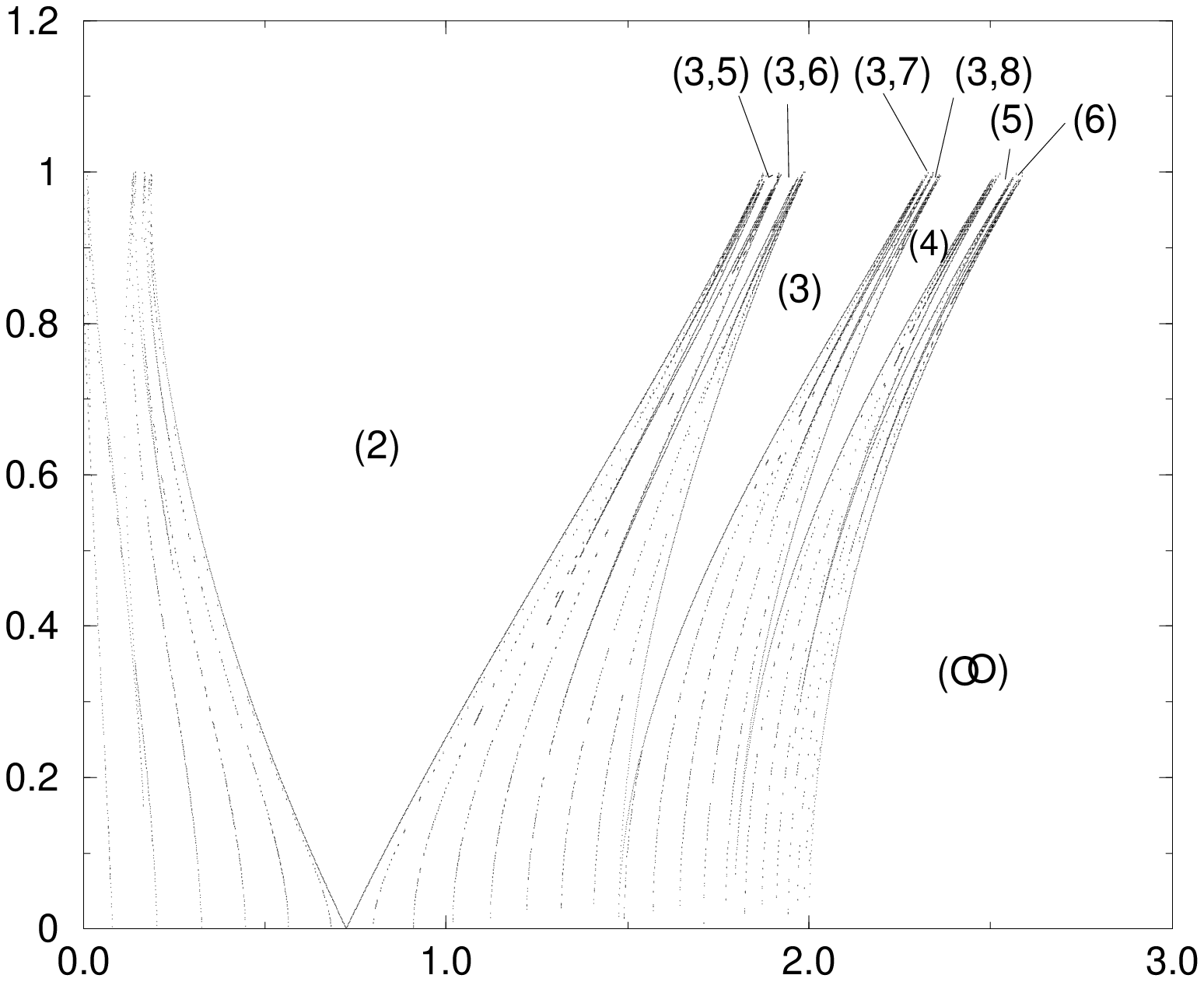}}
\put(-2,30){$\epsilon$}
\put(40,1){$E$}
\end{picture}
\caption{ Phase diagram of the Harper map for $\omega=\omega^*$. The
labels of the L phases are their winding numbers. The L-E transitions,
represented as the boundaries of the L phases, end in SNAs for
$\epsilon = 1$. Unlike the phase diagrams for rational $\omega$, there
are no co-stability regions and within each Arnol'd tongue there is
only one attractor. The tongues to the left of
the largest has variable winding number. Here we define $(k_1,k_2,k_3) \equiv
{\omega^*}^{k_1}+{\omega^*}^{k_2}+{\omega^*}^{k_3}$.}
\label{fig:phase-golden} 
\end{figure}

The number of phase transition lines increases as we approach $\omega^{*}$,
becoming infinite in the limit $\omega\to\omega^{*}$.
The phase diagram of the Harper
map for $\omega=\omega^*$ is in Fig.~\ref{fig:phase-golden}. The
labels within the Arnol'd tongues are the winding numbers of the
Localized phases.
 The L-E
transitions do not depend on the initial condition, then there are no
co-stability regions. There is only one
attractor for a given point inside a L phase, as can be checked by calculating the Lyapunov exponent.
This calculation also shows the fractal character of this phase
diagram~\cite{Bellisard}.
From now on a winding number given by a sum of powers of $\omega^{*}$,
say ${\omega^*}^{k_1}+{\omega^*}^{k_2}+{\omega^*}^{k_3}$, will be
represented by $(k_1,k_2,k_3)$.

\begin{figure} 
\narrowtext
\begin{picture}(73,33)(0,0)
 \put(5,-63){\epsfxsize=6in\epsffile{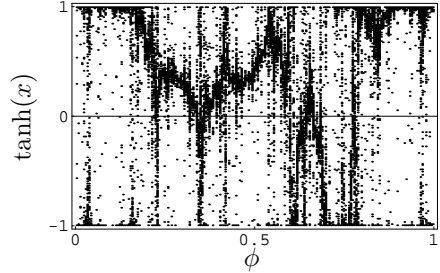}}
 \put(41,1){$\phi$}
 \put(10,15){\rotatebox{90}{$\tanh(x)$}}
\end{picture}
\caption{ Phase portrait of the SNA for $\omega=\omega^*$ and $\epsilon=1$
at $E=2.3447775\ldots$, the endpoint of the left boundary of the
Localized phase with winding number ${\omega^*}^4$ in
Fig.~\ref{fig:phase-golden}.}
\label{fig:sna} 
\end{figure}

\begin{figure} 
\narrowtext
\begin{picture}(73,33)(0,0)
 \put(5,-3){\epsfysize=1.6in\epsfxsize=2.45in\epsffile{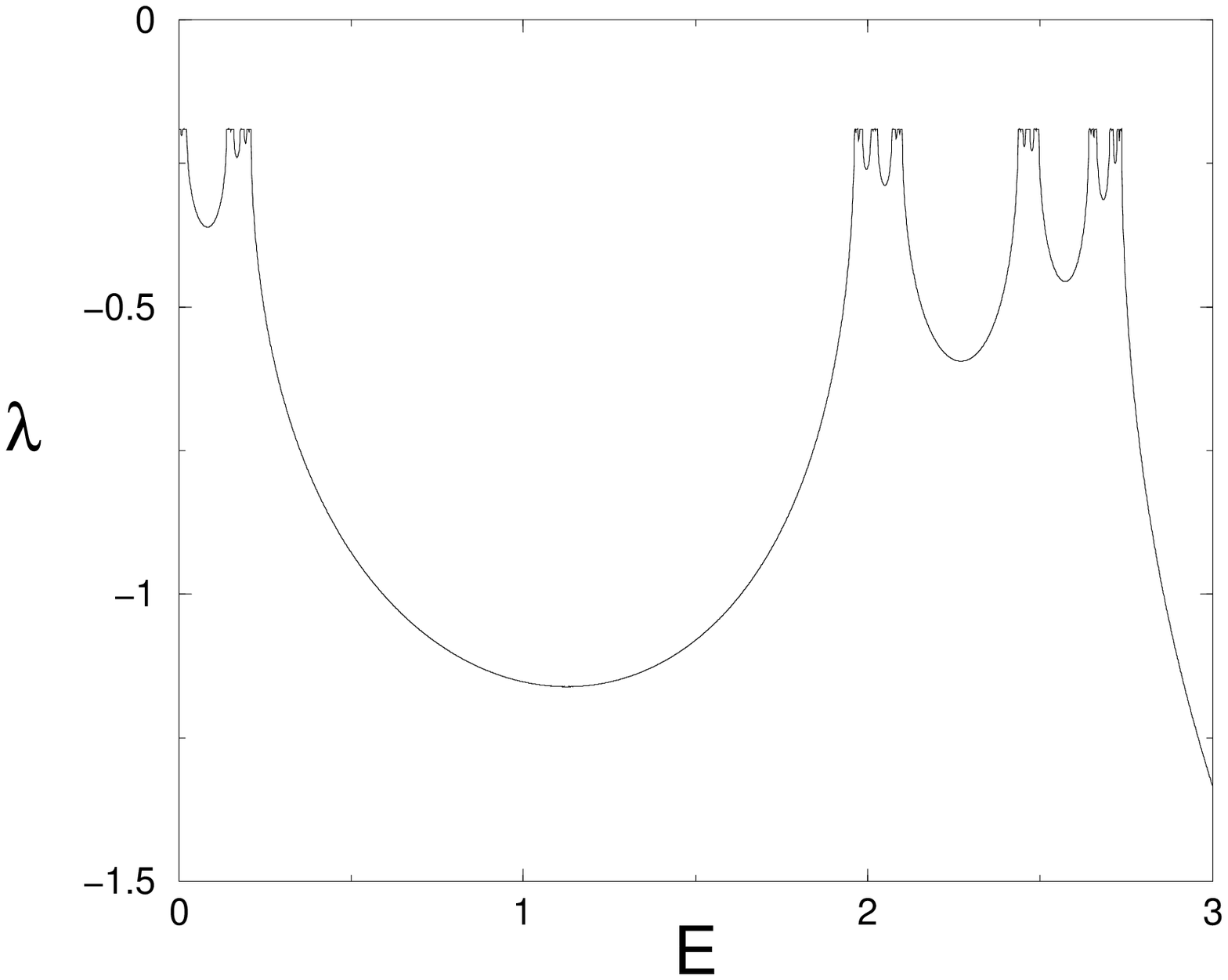}}
\end{picture}
\caption{Lyapunov exponent as a function of the energy,
for $\epsilon = 1.1$ and $ \omega=\omega^*$. The L phases occupy all the
line, since $\lambda <0$.}
\label{fig:lya-overcrit} 
\end{figure}

\begin{figure} 
\narrowtext
\begin{picture}(73,26)(0,0)
 \put(-5,-2){\epsfxsize=3.5in\epsffile{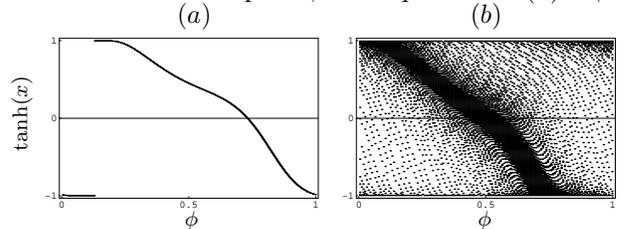}}
 \put(21,1){\footnotesize$\phi$}
 \put(60,1){\footnotesize$\phi$}
 \put(-2,11){\footnotesize\rotatebox{90}{$\tanh(x)$}}
 \put(20,28){$(a)$}
 \put(59,28){$(b)$}
\end{picture}
\caption{Typical Localized and Extended configurations for
$\omega=\omega^*$ and $\epsilon=0.2$. (a)~$E=0.938$. 
(b)~$E=0.94$. The {\em locus} of the L-E transition is $E=0.9391\ldots$.}
\label{fig:LE-golden} 
\end{figure}

In the case of Localized phases, the winding numbers for
$\omega=\omega^*$ are limit values of those
for $\omega_m=F_{m-1}/F_m$. In general,
the phase with winding number $\Omega={w^*}^k=(k)$ ($k$ integer) is the 
limit of the phases with
$\Omega=F_{m-k}/F_m$, when $m\rightarrow \infty$. Moreover, the 
L-E phase transitions end at the line
$\epsilon=1$. Their endpoints have $\lambda=0$ and
are strange nonchaotic attractors 
like the one in $(\epsilon,E)=(1,0)$, thoroughly analyzed in
reference~\cite{Prasad}. Therefore, there is evidence of the occurrence
of an infinite number of SNAs in this phase
diagram. One of them is shown in Fig.~\ref{fig:sna}.
For $\epsilon>1$ all Lyapunov exponents become 
negative although they maintain their \textit{bumpy} structure (see
Fig.~\ref{fig:lya-overcrit}). This means that the transition between
different L regions is smooth without crossing an E region. This 
resembles the liquid-vapor phase transition of the water. Above 
the critical point the change between liquid and vapor is smooth. 
The typical E and L attractors are also different for
$\omega=\omega^*$. In this case, the L-E transition is
between a unidimensional and a two-dimensional attractor,
as displayed in Fig.~\ref{fig:LE-golden}. The change in
the energy and/or $\epsilon$ can cause a spread in the unidimensional
L attractor towards
a two-dimensional E phase, in the plane $\tanh(x) \times \phi$. 

\begin{figure} 
\narrowtext
\begin{picture}(73,92)(0,0)
 \put(-1.7,46){\epsfxsize=3.17in\epsfysize=2in\epsffile{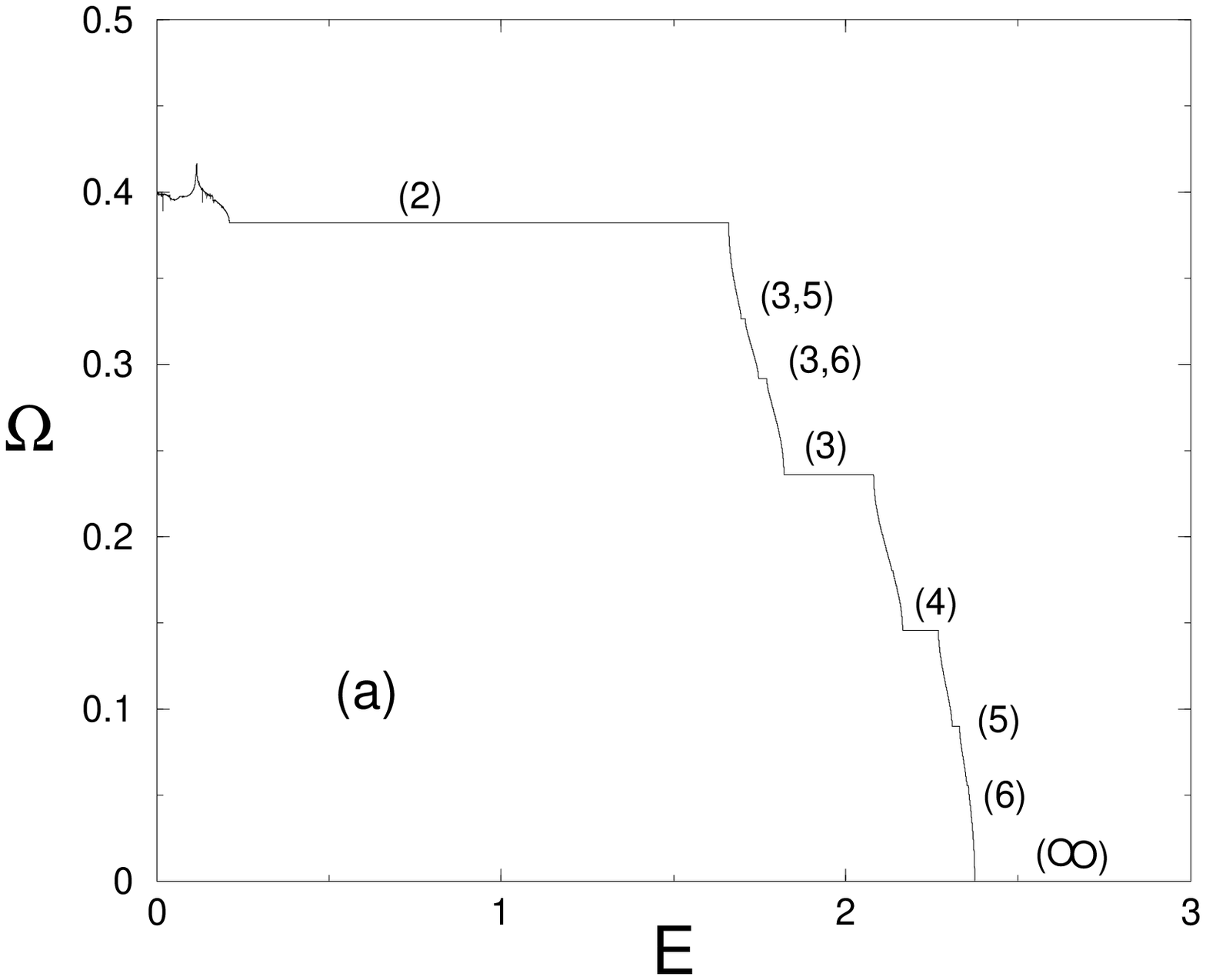}}
 \put(-3,-2){\epsfxsize=3.2in\epsfysize=2in\epsffile{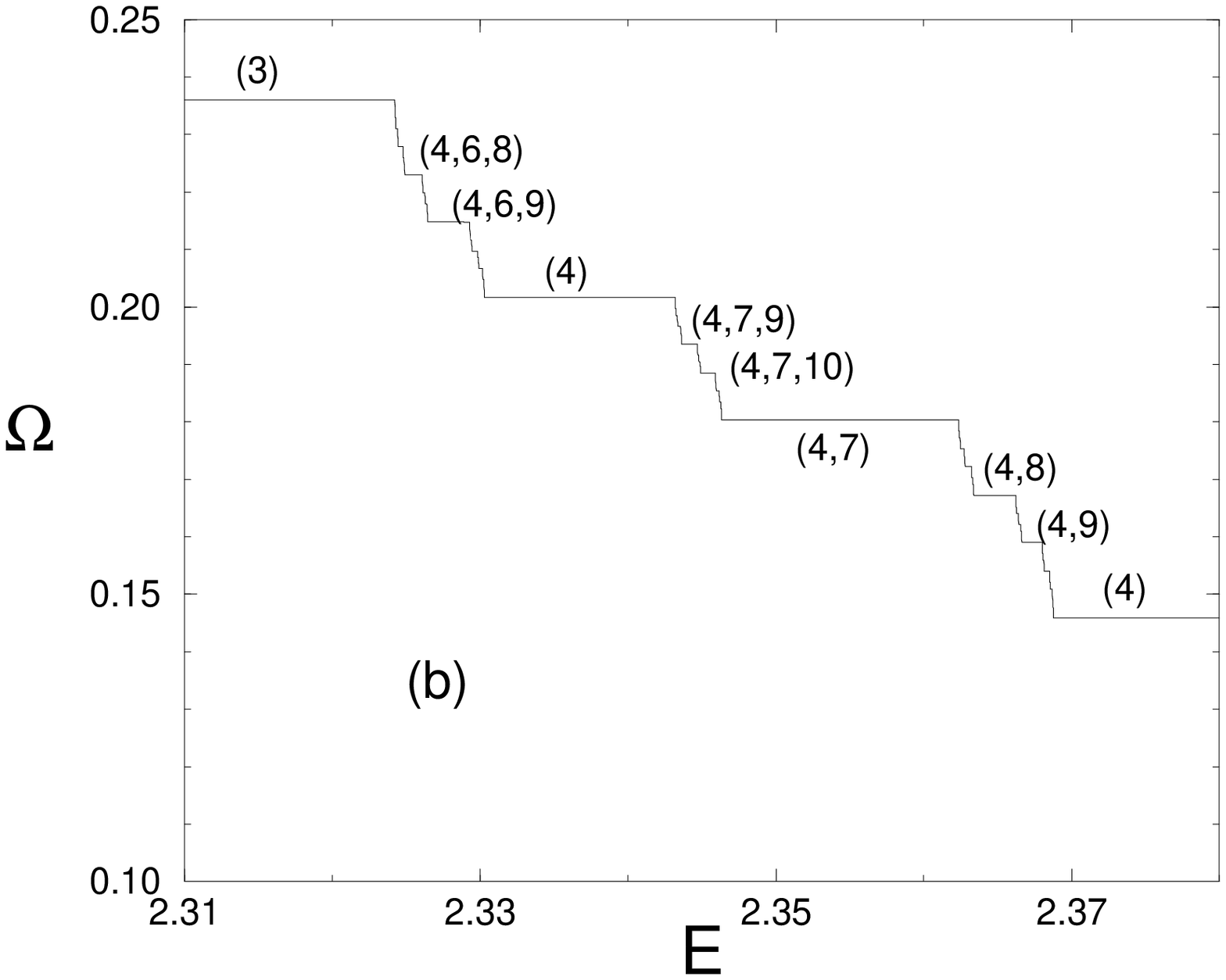}}
\end{picture}
\caption{ (a) Winding number as a function of energy, for
$\omega=\omega^*$ and $\epsilon = 0.8$. The plateaux correspond to L
phases and their winding numbers are displayed. They are sums of
powers of the golden mean. Between two plateaux there are other sets of
plateaux, displaying a fractal structure. (b) Detail of the golden
staircase for  $\epsilon = 1$.}
\label{fig:wind-golden} 
\end{figure}

One of the most original features of the phase diagram for
$\omega=\omega^*$ is the {\em irrationality of the
winding numbers of the Localized phases within the Arnol'd tongues}.
They are sums of powers of the golden mean. This is the first phase diagram to
show it, as far as we know. Starting from the left to
the right, the largest phases have winding numbers
such that $\Omega={\omega^*}^j$,
where $j=2,3,4,\ldots,\infty$. The plot of $\Omega$ vs. energy 
for $\epsilon=0.8$ shows a number of
plateaux, corresponding to L phases. We name it {\bf Golden Staircase}
(see Fig.~\ref{fig:wind-golden})
and its structure is completely different from the Farey tree
observed in many modulated and dynamical systems~\cite{KT}.

There is an order beneath the irrational labels of the L phases. Firstly,
the main sequence of integer powers of the golden mean $(k)$ , for
$k=2,3,4,\ldots$ occurs in a similar way between two elements of the 
same sequence. For example, take $k=3$ and $k=4$. These labels can be rewritten as
$(3)=(4,5)$ and $(4)=(4,\infty)$.
Between them there are the winding numbers $(4,j)$, for
$j=6,7,8,\ldots$. Secondly, given two neighboring tongues labelled
by, say, $(k)$ and $(k+1)$ there
is a rule to find out which are the two largest tongues between them. For
example take the two largest tongues
between $(3)$ and $(4)$: $(4,6)$ and
$(4,7)$. They correspond to the lowest powers of
$\omega^*$ to be summed up to $(4)$. It is now easy to go one step down
in the hierarchy; take  the powers $(4,6)$ and $(4,7)$, which can be
rewritten respectively, as  $(4,7,8)$ and $(4,7,\infty)$. Between them
we find the winding numbers $(4,7,j)$, for $j=9,10,11,
\ldots$, where $(4,7,9)$ and $(4,7,10)$ are connected to the largest
tongues. The other
steps can be deducted following the same reasoning.
\begin{figure} 
\narrowtext
\begin{picture}(73,28)(0,0)
 \put(-3,0){\epsfxsize=3.26in\epsfysize=1.2in\epsffile{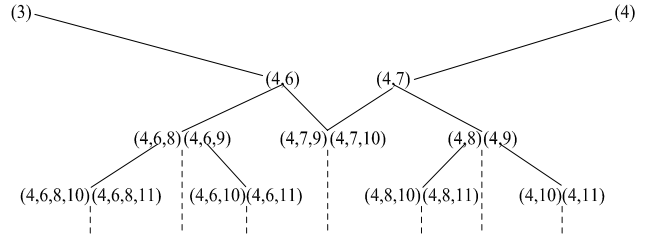}}
\end{picture}
\caption{Partial hierarchy of the L phases.}
\label{fig:hierarchy} 
\end{figure}
We conjecture that the lock-in of phases with irrational winding 
numbers should be a common feature for other
irrational values of $\omega$ for the Harper map~\cite{YL_1}. Besides, 
the structure of the diagram $\Omega$ vs. energy should resemble the
one of the Golden Staircase. 

In this letter we have presented phase diagrams for the Harper map for
$\omega=5/8$, $8/13$ and $\omega^*$. We described in detail the
occurrence of finite or infinite number of Arnol'd tongues,
strange nonchaotic  attractors and the L-E transition in each case.
We also reported the occurrence of co-stability
regions and the peculiar structure of the staircases representing the
winding number.

The study of the density of states with irrational winding numbers,
following Thouless~\cite{T_1} will be subject of forthcoming publication.

The work of PCF is supported by PRAXIS XXI/BD/11461/97 grant from
FCT (Portugal).
The work of FPM is supported by EPRSC studentship 97304299 and by
Fondazione A. Della Riccia. MHRT thanks O. Kinouchi for helpful discussions.

\end{multicols}

\begin{thebibliography}{99}
\bibitem{winding} We define winding number as $\Omega = \lim_{N \to
\infty} S/(2N)$, where $S$ is the number of times the variable $x$ changes
sign during $N$ iterations.
\end{thebibliography}
\end{document}